\documentclass[prc,twocolumn,nofootinbib,superscriptaddress,aps,amsmath,amssymb]{revtex4-1}

\usepackage{graphicx}
\usepackage{dcolumn}
\usepackage{bm}
\usepackage{hyperref}
\usepackage{tabularx}
\newcolumntype{Y}{>{\centering\arraybackslash}X}
\usepackage{xspace}
\makeatletter
\DeclareRobustCommand\onedot{\futurelet\@let@token\@onedot}
\def\@onedot{\ifx\@let@token.\else.\null\fi\xspace}

\newcommand{\etal}{{\it et al}\onedot}

\newcommand{\eg}{{e.g}\onedot}

\makeatother

\newcommand{\fmi}{\ensuremath{\,\text{fm}^{-1}}}

\newcommand{\MeV}{\ensuremath{\,\text{MeV}}}

\newcommand{\GeVi}{\ensuremath{\,\text{GeV}^{-1}}}

\newcommand{\MeVdivA}{\ensuremath{\,\text{MeV/A}}}

\newcommand{\eMax}{\ensuremath{e_{\text{Max}}}}
\newcommand{\EMax}{\ensuremath{E_{3\text{Max}}}}
\newcommand{\Sn}{\ensuremath{S_{2\text{n}}}}

\newcommand{\op}[1]{\ensuremath{#1}}
\newcommand{\rOV}{\ensuremath{\vec{\op{r}}}}
\newcommand{\ROV}{\ensuremath{\vec{\op{R}}}}

\newcommand\Tstrut{\rule{0pt}{2.6ex}}         
\newcommand\Bstrut{\rule[-0.9ex]{0pt}{0pt}}   

\begin{document}

\title{Saturation with chiral interactions and consequences for finite nuclei}

\author{J.~Simonis}
\email[E-mail:~]{simonis@theorie.ikp.physik.tu-darmstadt.de}
\affiliation{Institut f\"ur Kernphysik, Technische Universit\"at Darmstadt, 64289 Darmstadt, Germany}
\affiliation{ExtreMe Matter Institute EMMI, GSI Helmholtzzentrum f\"ur Schwerionenforschung GmbH, 64291 Darmstadt, Germany}
\author{S.~R.~Stroberg}
\email[E-mail:~]{sstroberg@triumf.ca}
\affiliation{TRIUMF, 4004 Wesbrook Mall, Vancouver, BC V6T 2A3, Canada}
\author{K.~Hebeler}
\email[E-mail:~]{kai.hebeler@physik.tu-darmstadt.de}
\affiliation{Institut f\"ur Kernphysik, Technische Universit\"at Darmstadt, 64289 Darmstadt, Germany}
\affiliation{ExtreMe Matter Institute EMMI, GSI Helmholtzzentrum f\"ur Schwerionenforschung GmbH, 64291 Darmstadt, Germany}
\author{J.~D.~Holt}
\email[E-mail:~]{jholt@triumf.ca}
\affiliation{TRIUMF, 4004 Wesbrook Mall, Vancouver, BC V6T 2A3, Canada}
\author{A.~Schwenk}
\email[E-mail:~]{schwenk@physik.tu-darmstadt.de}
\affiliation{Institut f\"ur Kernphysik, Technische Universit\"at Darmstadt, 64289 Darmstadt, Germany}
\affiliation{ExtreMe Matter Institute EMMI, GSI Helmholtzzentrum f\"ur Schwerionenforschung GmbH, 64291 Darmstadt, Germany}
\affiliation{Max-Planck-Institut f\"ur Kernphysik, Saupfercheckweg 1, 69117 Heidelberg, Germany}

\begin{abstract}
We explore the impact of nuclear matter saturation on the properties
and systematics of finite nuclei across the nuclear chart.  Using the
ab initio in-medium similarity renormalization group (IM-SRG), we
study ground-state energies and charge radii of closed-shell nuclei
from $^4$He to $^{78}$Ni, based on a set of low-resolution two- and
three-nucleon interactions that predict realistic saturation
properties. We first investigate in detail the convergence properties
of these Hamiltonians with respect to model-space truncations for both
two- and three-body interactions. We find one particular interaction
that reproduces well the ground-state energies of all closed-shell nuclei
studied. As expected from their saturation points relative to this
interaction, the other Hamiltonians underbind nuclei, but lead to a
remarkably similar systematics of ground-state energies.  Extending
our calculations to complete isotopic chains in the $sd$ and $pf$
shells with the valence-space IM-SRG, the same interaction reproduces
not only experimental ground states but two-neutron-separation energies
and first excited $2^+$ states. We also calculate radii with the
valence-space IM-SRG for the first time. Since this particular
interaction saturates at too high density, charge radii are still too
small compared with experiment. Except for this underprediction, the
radii systematics is, however, well reproduced.  Our results highlight
the importance of nuclear matter as a theoretical benchmark for the
development of next-generation chiral interactions.
\end{abstract}

\maketitle

\section{Introduction}
\label{sec:Intro}

A central goal of nuclear theory is an accurate ab initio description
of nuclei from the valley of stability to the limits of existence
based on a common Hamiltonian including theoretical uncertainties.
Recent progress in chiral effective field theory
(EFT)~\cite{Epel09RMP,Mach11PR} and renormalization group
methods~\cite{Bogn10PPNP,Furn13RPP} methods provide a framework
for addressing this goal.  The enhanced
convergence properties of the resulting interactions and
methodological advances in many-body theory, \eg, with coupled-cluster
(CC) theory~\cite{Hage14RPP}, self-consistent Green's function
methods~\cite{Soma14GGF2N3N}, or the in-medium similarity
renormalization group (IM-SRG)~\cite{Herg16PR}, which exhibit a
polynomial scaling in the mass number, have increased the reach of ab
initio calculations to the medium-mass region or beyond. While these
large-space methods are still limited to closed-shell and neighboring
nuclei, or to even-even systems, doubly open-shell nuclei can be
accessed by ab initio valence-space
methods~\cite{Bogn14SM,Jans14SM,Jans16SM,Stro16TNO,Stro17ENO}.

These developments have enabled a clear demonstration of the
importance of three-nucleon (3N) forces for understanding the
structure of medium-mass nuclei~\cite{Hamm13RMP,Hebe15ARNPS} and for
realistic saturation properties of nuclear
matter~\cite{Bogn05nuclmat,Hebe11fits,Hage14ccnm,Carb13nm,Cora14nmat}.
Indeed, the role of 3N forces in saturation was suggested long
ago~\cite{Whee37molec,Drel53sat}, but difficulties in formulating
consistent 3N forces and solving the resulting many-body problem
hindered progress.  More recently, the difficulty has been in
constructing a chiral interaction which simultaneously reproduces
saturation, few-body observables, and spectroscopy.  In
Ref.~\cite{Roth14SRG3N}, a good description of ground-state energies
in the region of $^{16}$O was obtained using a chiral two- and
three-nucleon interaction. However, this interaction yields radii
which are too small~\cite{Cipo13Ox}, and increasingly severe
overbinding for heavier nuclei~\cite{Bind14CCheavy}. An alternative
approach~\cite{Ekst15sat}, fitting directly to some medium-mass nuclei
successfully reproduced saturation, but two-nucleon (NN) scattering
data were only fit up to $35 \MeV$.

In this work, we investigate ground-state energies and charge radii of
finite nuclei based on chiral low-resolution NN and 3N interactions
with realistic saturation properties, which also reproduce scattering
data with high precision.  We use the closed-shell and valence-space
formulation of the IM-SRG. This enables a unique and broad access
across the nuclear chart. In Sec.~\ref{subsec:Interactions}, we
describe the Hamiltonians used in this work, while
Sec.~\ref{subsec:IMSRG} gives a short overview on the IM-SRG formalism
for closed-shell nuclei and the decoupling of valence-space
interactions for open-shell nuclei. The model-space convergence of
ground-state energies and charge radii for different resolution scales
and low-energy couplings is studied in detail in
Sec.~\ref{sec:ClosedShell}.  Finally, Sec.~\ref{sec:Chains} presents
valence-space results for different $sd$- and $pf$-shell isotopic
chains, including first results for radii in the valence-space IM-SRG.

\subsection{Chiral interactions and saturation}
\label{subsec:Interactions}

At the NN level, we start from the next-to-next-to-next-to-leading
order (N$^3$LO) $500\MeV$ potential of Entem and Machleidt
(EM)~\cite{Ente03EMN3LO}. We then use the similarity renormalization
group (SRG)~\cite{Bogn07SRG,Bogn10PPNP} to evolve this interaction to
a series of low-resolution scales $\lambda_{\rm NN} = 1.8, 2.0, 2.2
\fmi$. Taking chiral EFT as a general low-momentum basis and assuming
the long-range couplings $c_i$ to be invariant under the SRG
transformation, we combine each SRG-evolved NN interaction with the
leading N$^2$LO 3N forces~\cite{Kolc94fewbody,Epel02fewbody}, where
the $c_i$ couplings in the two-pion-exchange 3N interaction are taken
consistently with the NN interaction: $c_1 = -0.81 \GeVi$, $c_3 = -3.2
\GeVi$, $c_4 = 5.4\GeVi$. In addition, to probe uncertainties in the
$c_i$ couplings, we use 3N forces with the $c_i$ values obtained from
the Nijmegen NN partial-wave analysis (PWA): $c_1 = -0.76 \GeVi$, $c_3
=-4.78 \GeVi$, $c_4 = 3.96 \GeVi$~\cite{Rent03ciPWA} for the
$\lambda_{\rm NN} = 2.0 \fmi$ interaction. For all Hamiltonians, the
low-energy couplings $c_D$, $c_E$ in the 3N one-pion-exchange and 3N
contact interaction have been fit to the $^3$H binding energy and
$^4$He charge radius using Faddeev and Faddeev-Yakubovsky
calculations~\cite{Hebe11fits} with a nonlocal 3N regulator and cutoff
$\Lambda_{\rm 3N} = 2.0 \fmi$.

These chiral NN+3N Hamiltonians were first employed to study
symmetric~\cite{Hebe11fits} and, more recently, also asymmetric
nuclear matter~\cite{Dris14asymmat,Dris16asym}.  The first application
to finite nuclei was in a valence-space study of $sd$-shell
isotopes~\cite{Simo16unc} and in coupled-cluster calculations of
selected Ca~\cite{Hage16NatPhys,Ruiz16Calcium} and Ni
isotopes~\cite{Hage16Ni78}.  Of particular importance to this work is
that in symmetric nuclear matter the $\lambda_{\text{NN}}/
\Lambda_{\text{3N}} = 1.8/2.0$ (EM) interaction yields an energy per
particle in good agreement with the empirical value (at saturation
density with a Hartree-Fock spectrum slighty too
bound~\cite{Dris16asym}), although at a somewhat too high density. The
other interactions $2.0/2.0$ (EM), $2.2/2.0$ (EM), $2.0/2.0$ (PWA)
saturate at decreasingly smaller energy and density~\cite{Dris16asym}.

\subsection{In-medium SRG}
\label{subsec:IMSRG}

In the IM-SRG approach~\cite{Herg16PR}, starting from the full NN+3N
Hamiltonian, we first solve the Hartree-Fock equations to obtain a
suitable reference state.  We then rewrite the Hamiltonian in
normal-ordered form:
\begin{align}
H &= E_0 +  \sum_{ij}f_{ij}\{a_i^{\dagger}a_j\}
+ \frac{1}{4}\sum_{ijkl}\Gamma_{ijkl}\{a^{\dagger}_ia^{\dagger}_ja_la_k\} \nonumber \\
&\quad + \frac{1}{36}\sum_{ijklmn}W_{ijklmn} \{a^{\dagger}_ia^{\dagger}_ja^{\dagger}_ka_na_ma_l\} \,,
\end{align}
where the braces denote a string of creation and annihilation
operators normal ordered with respect to the reference state, and the
resulting in-medium zero-, one-, and two-body operators, $E_0$, $f$,
and $\Gamma$, respectively, represent the starting values for the
IM-SRG flow equations.  In the normal-ordered two-body approximation,
in which we work, the residual three-body term $W$ is discarded.  We
use the Magnus formulation presented in Ref.~\cite{Morr15Magnus} to
generate an explicit unitary transformation that decouples the
reference state from excitations.  This transformation can
subsequently be applied to any operator, in particular the radius
operator discussed below.  For calculations of open-shell nuclei, we
use the valence-space formulation of the IM-SRG
(VS-IM-SRG)~\cite{Tsuk12SM,Bogn14SM,Stro16TNO} based on the ensemble
normal-ordering discussed in Ref.~\cite{Stro17ENO}, which captures the
bulk effects of residual 3N forces among valence nucleons.  A
valence-space Hamiltonian is then produced specifically for each
nucleus, which is diagonalized with the NuShellX shell-model
code~\cite{Brow14NuShellX} to obtain ground- and excited-state
energies in the valence space.

\section{Closed-shell nuclei}
\label{sec:ClosedShell}

In this section, we analyze the model-space convergence of
closed-shell nuclei based on the four chiral low-resolution NN+3N
interactions introduced in Sec.~\ref{subsec:Interactions}.  In the
calculations we employ an angular-momentum-coupled basis built from
single-particle spherical harmonic-oscilator (HO) states with quantum
numbers $e=2n+l \leqslant \eMax$.  We employ partial-wave decomposed
3N matrix elements in a Jacobi-momentum basis and include partial
waves up to the total three-body angular momentum $\mathcal{J}
\leqslant 9/2$.  Furthermore in order to manage computational storage
requirements, we introduce an additional cut
$e_{1}+e_{2}+e_{3}\leqslant \EMax < 3 \eMax$ for 3N matrix elements.
For the analysis of the convergence behavior presented below, we study
$\eMax/\EMax=10/14, 12/14, 14/14, 14/16$, and $14/18$.

\begin{figure*}[t]
\includegraphics[width=\textwidth]{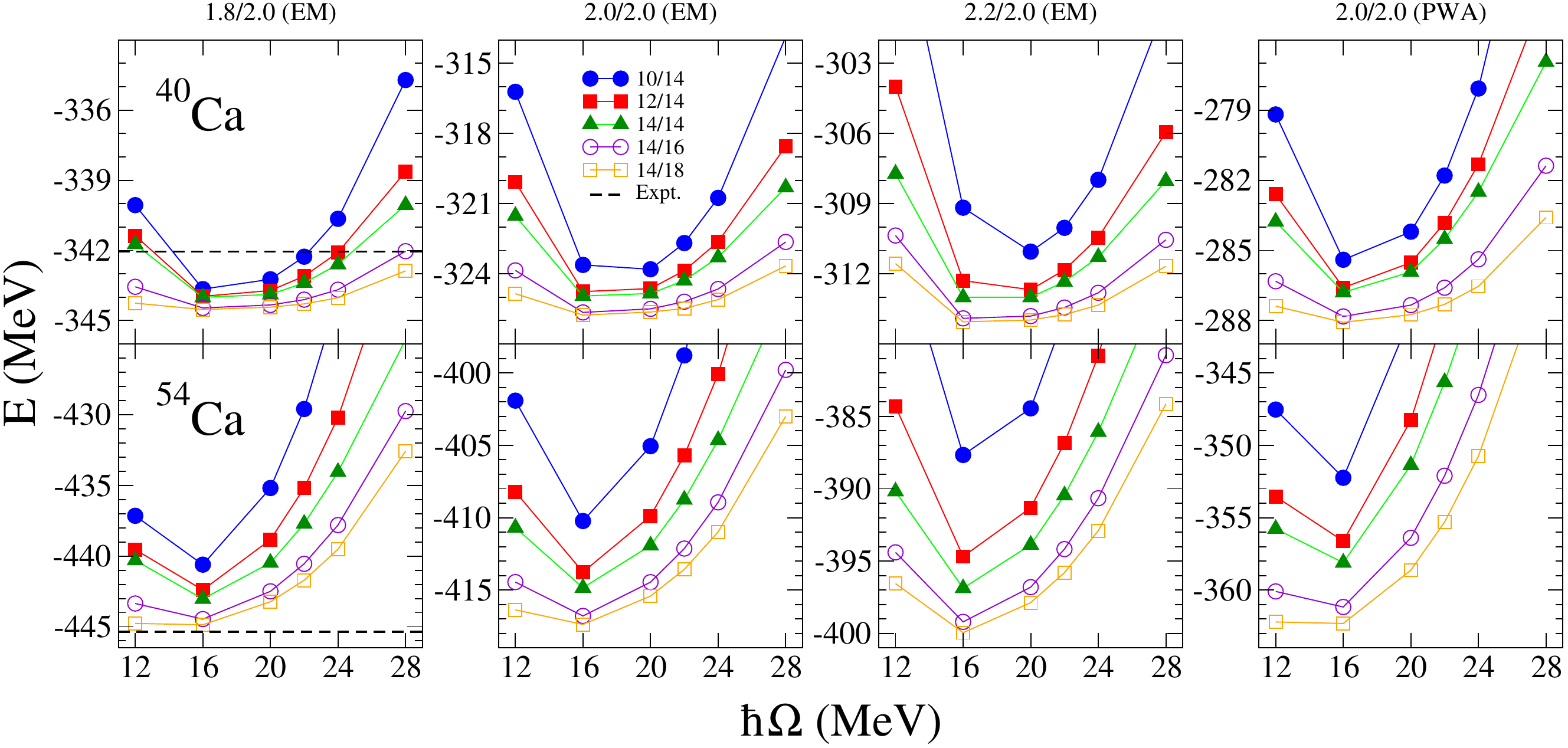}
\caption{Convergence of ground-state energies for $^{40}$Ca (top panels) 
and $^{54}$Ca (bottom panels) calculated with the closed-shell
IM-SRG. The column heading specifies the input Hamiltonian:
$\lambda_{\text{NN}}/ \Lambda_{\text{3N}} = 1.8/2.0$ (EM) (left),
2.0/2.0 (EM) (middle-left), 2.2/2.0 (EM) (middle-right), and 2.0/2.0
(PWA) (right). In each panel, we show results obtained for
harmonic-oscillator frequencies $\hbar\Omega=(12-28)\MeV$ and
different truncations of the single-particle basis $\eMax/\EMax=10/14$
(filled circles), $12/14$ (filled squares), $14/14$ (filled
triangles), $14/16$ (empty circles), and $14/18$ (empty squares). The
experimental ground-state energies from the atomic mass evaluation
(AME) 2012~\cite{Wang12AME12} are given by the dashed lines.}
\label{fig:CaEgsConvergence}
\end{figure*}

\begin{table*}[t]
\centering
\begin{tabularx}{\textwidth}{l Y Y Y Y | Y Y Y Y}
\hline
\hline
& \multicolumn{4}{c|}{$^{54}$Ca} & \multicolumn{4}{c}{$^{78}$Ni} \Tstrut\\
Hamiltonian & (10$\rightarrow$12)/14 & (12$\rightarrow$14)/14 & 14/(14$\rightarrow$16) & 14/(16$\rightarrow$18) & (10$\rightarrow$12)/14 & (12$\rightarrow$14)/14 & 14/(14$\rightarrow$16) & 14/(16$\rightarrow$18) \Tstrut\\[1mm]
\hline
1.8/2.0 (EM) & 1.8 & 0.6 & 1.4 & 0.4 & 3.3 & 0.9 & 4.4 & 2.0 \Tstrut\\
2.0/2.0 (EM) & 3.5 & 1.1 & 1.9 & 0.6 & 7.4 & 2.1 & 5.7 & 2.8 \Tstrut\\
2.2/2.0 (EM) & 7.0 & 2.2 & 2.3 & 0.7 & 14.7 & 5.0 & 6.6 & 3.4 \Tstrut\\
2.0/2.0 (PWA)& 4.4 & 1.5 & 3.1 & 1.1 & 9.2 & 3.0 & 9.6 & 5.0 \Tstrut\Bstrut\\
\hline
\hline
\end{tabularx}
\caption{Convergence of ground-state energies of $^{54}$Ca and $^{78}$Ni 
for the four Hamiltonians considered. The table lists the change in the
ground-state energy when increasing $\eMax \to \eMax+2$ ($\EMax \to 
\EMax+2$) at fixed $\EMax$ ($\eMax$) for harmonic-oscillator frequency
$\hbar\Omega=16\MeV$.}
\label{tab:egs_convergence}
\end{table*}

In addition to ground-state energies, we also investigate the
convergence behavior of charge radii. These results are obtained by
normal-ordering and evolving the intrinsic proton mean-square radius
operator (see, e.g., Ref.~\cite{Kamu97rmsradii}),
\begin{equation}
R^2_{p} = \frac{1}{Z} \sum\limits_i^Z \left( \rOV_i - \ROV \right)^2 \,,
\label{eq:def_rp}
\end{equation}
where $i$ runs over the proton coordinates, and $\ROV$ is the
center-of-mass coordinate. Note that the proton mean-square radius operator is
not free-space SRG evolved, consistent with the determination of the 3N couplings to
the charge radius of $^4$He. We obtain charge radii by applying the
corrections arising from the mean-square charge radii of the proton
and the neutron as well as the relativistic Darwin-Foldy and
spin-orbit corrections:
\begin{equation}
R_\text{ch} = \sqrt{R^2_p + \left\langle r^2_p \right\rangle + \frac{N}{Z}\left\langle r^2_n \right\rangle + \frac{3}{4 M^2_p c^4} + \left\langle r^2 \right\rangle_{\rm so}} \,,
\label{eq:def_rch}
\end{equation}
with values of $\left\langle r^2_p \right\rangle$ and $\left\langle
r^2_n \right\rangle$ taken from Ref.~\cite{PDG16review}.  The spin-orbit
correction~\cite{Ong10spinorbit} is calculated from
\begin{equation}
\langle r^2 \rangle_{\rm so} = \frac{1}{Z} \sum_{i=1}^{A} \langle r_i^2 \rangle_{\rm so} = - \frac{1}{Z} \sum_i \frac{\mu_i}{M^2}(\kappa_i+1) \,,
\end{equation}
with the magnetic moments of the proton, $\mu_{p}=2.793 \mu_{N}$, and
the neutron, $\mu_{n}=-1.913 \mu_{N}$, and the definition
\begin{equation}
\kappa = \begin{cases} l \,, & j=l-\frac{1}{2} \\ -(l+1) \,, & j=l+\frac{1}{2} \,. \end{cases}
\end{equation}
Additionally, two-body currents, due to the coupling of the photon to
pions and to two nucleons, contribute to the charge radius, but this
correction is neglected here.

\begin{figure*}[t]
\includegraphics[width=2\columnwidth]{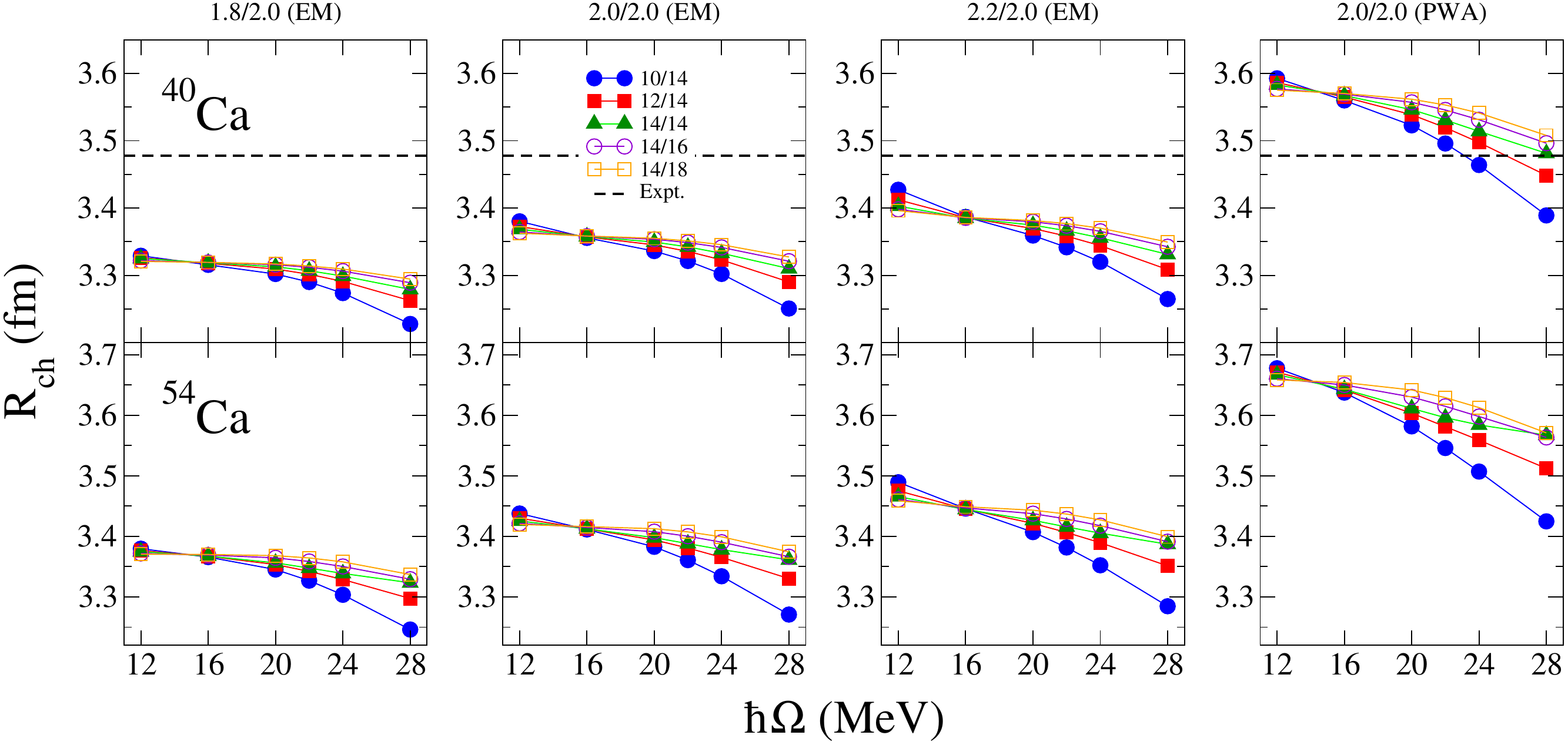}
\caption{Convergence of charge radii for $^{40}$Ca (top panels) and 
$^{54}$Ca (bottom panels) calculated with the closed-shell IM-SRG. The
legend is as in Fig.~\ref{fig:CaEgsConvergence}. The experimental charge
radius for $^{40}$Ca~\cite{Ange13rch} is given by the dashed line.}
\label{fig:CaRchConvergence}
\end{figure*}

In Fig.~\ref{fig:CaEgsConvergence}, we show the model-space
convergence for ground-state energies of $^{40}$Ca and $^{54}$Ca.  The
energy minima are almost independent of the four different NN+3N
interactions, typically located near $\hbar\Omega=16\MeV$.  While the
ground-state energy of $^{40}$Ca is well converged for the different
Hamiltonians, in $^{54}$Ca convergence from $\eMax/\EMax = 10/14$ to
14/14 is only obtained for the interactions with lower resolution
scales $\lambda_{\rm NN} = 1.8$ and $2.0 \fmi$. In
Table~\ref{tab:egs_convergence}, we list the change in the
ground-state energy with increasing $\eMax/\EMax$. We clearly see the
1.8/2.0 (EM) interaction is better converged from $\eMax/\EMax=12/14$
to 14/14, where the energy decreases by only $0.6\MeV$ total, compared
with the 2.2/2.0 (EM) interaction, where this decrease is $2.2\MeV$.

In addition, we investigate the impact of increasing the 3N cut
$\EMax=14\rightarrow 18$ for $\eMax=14$ both in
Fig.~\ref{fig:CaEgsConvergence} and Table~\ref{tab:egs_convergence}.
In the case of the 1.8/2.0 (EM) interaction, the ground-state energy
of $^{54}$Ca decreases by $0.4\MeV$ for $\EMax=16 \rightarrow 18$,
while this decrease of $0.7\MeV$ is only slightly larger for the
2.2/2.0 (EM) interaction, indicating both are relatively well
converged in terms of $\EMax$.  The largest impact is seen with the
2.0/2.0 (PWA) interaction, where the difference amounts to $1.1\MeV$.
While the ground-state energies calculated with the 1.8/2.0 (EM) interaction agree with
experiment to $\approx 1\%$, the other three Hamiltonians
predict energies that are significantly underbound.

In Fig.~\ref{fig:CaRchConvergence}, we show the model-space
convergence of the charge radii for $^{40}$Ca and $^{54}$Ca.  Although
the ground-state energy of $^{40}$Ca calculated from the 1.8/2.0 (EM)
interaction is in remarkable agreement with experiment, the
corresponding charge radius, shown in the left column of
Fig.~\ref{fig:CaRchConvergence}, is significantly smaller than
experiment.  With increasing SRG resolution scale $\lambda_{\rm NN}$,
the charge radii increase but are still too small compared to
experiment, while for the 2.0/2.0 (PWA) Hamiltonian, the calculated
charge radius is instead somewhat too large.  It will be very
interesting to compare the charge-radii calculations for $^{54}$Ca,
shown in the lower panels of Fig.~\ref{fig:CaRchConvergence} with
future experimental results. Even the recent measurement of the charge
radius of $^{52}$Ca~\cite{Ruiz16Calcium}, manifesting a strong
increase from $^{48}$Ca onward, could not be explained fully by
ab initio coupled-cluster calculations.

\begin{figure*}[t]
\includegraphics[width=2\columnwidth]{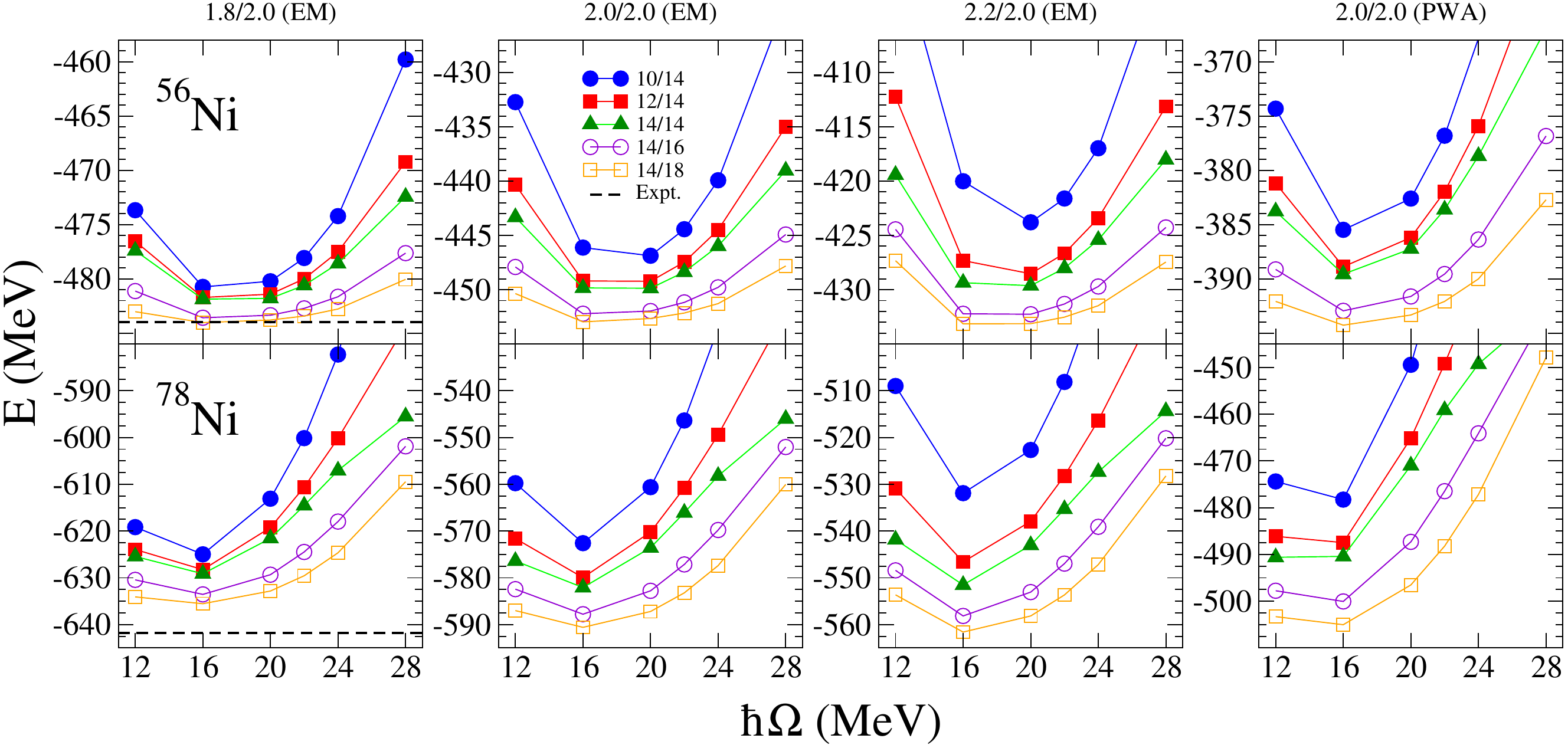}
\caption{Convergence of ground-state energies for $^{56}$Ni (top panels) 
and $^{78}$Ni (bottom panels) calculated with the closed-shell IM-SRG. 
The legend is as in Fig.~\ref{fig:CaEgsConvergence}. Note that the ground-state
energy for $^{78}$Ni is extrapolated.}
\label{fig:NiEgsConvergence}
\end{figure*}

\begin{figure*}[!t]
\includegraphics[width=2\columnwidth]{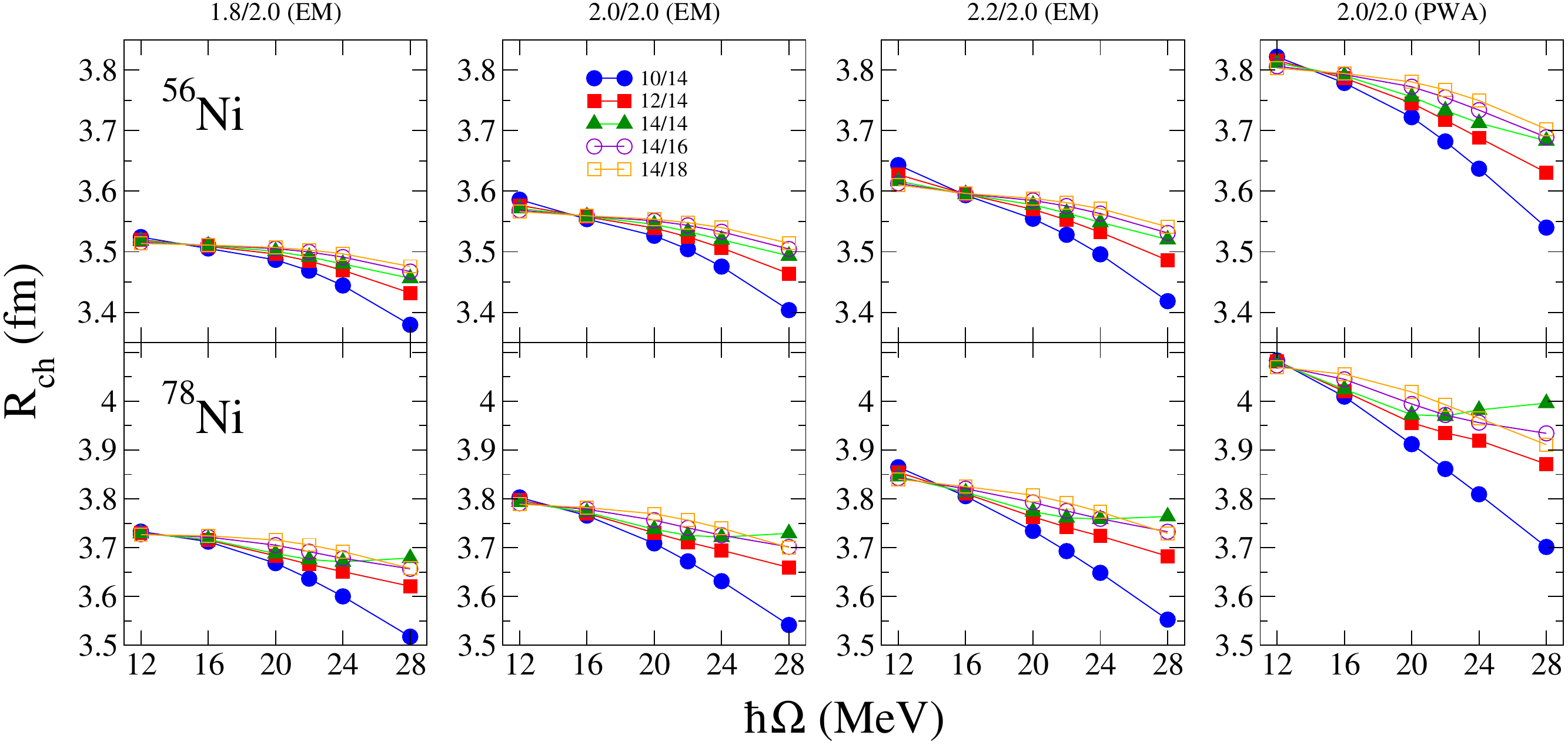}
\caption{Convergence of charge radii for $^{56}$Ni (top panels) and 
$^{78}$Ni (bottom panels) calculated with the closed-shell IM-SRG. The
legend is as in Fig.~\ref{fig:CaEgsConvergence}.}
\label{fig:NiRchConvergence}
\end{figure*}

In Fig.~\ref{fig:NiEgsConvergence}, we show the model-space
convergence for ground-state energies of $^{56}$Ni and $^{78}$Ni.
Similar to the calcium isotopes (see Fig.~\ref{fig:CaEgsConvergence})
the minima in the ground-state energies are near
$\hbar\Omega=16\MeV$. The energies obtained from the 1.8/2.0 (EM)
Hamiltonian are again in very good agreement with experiment, while
the other three Hamiltonians give results that are underbound to
different degrees.  The increase in particle number from calcium to
nickel clearly results in slower model-space convergence.  As seen in
Table~\ref{tab:egs_convergence}, enlarging the single-particle basis
from $\eMax=12 \rightarrow 14$ changes the ground-state energy of
$^{78}$Ni by $0.9\MeV$ for the 1.8/2.0 (EM) interaction at
$\hbar\Omega=16\MeV$, compared to $5.0\MeV$ for the 2.2/2.0 (EM)
interaction.  The change in energy when increasing the cut in the 3N
matrix elements from $\EMax = 16 \rightarrow 18$ is $2.0\MeV$ for
1.8/2.0 (EM), already not completely converged.  This effect is even
larger for 2.0/2.0 (EM) and 2.2/2.0 (EM) and maximal for 2.0/2.0
(PWA), where the change is $5.0 \MeV$.  Again, for the 1.8/2.0 (EM)
interaction, the agreement with experiment is good in both cases, but
it is clear that the model space must be increased beyond
$\eMax/\EMax=14/18$ to claim fully converged results in this region,
and likely for any nucleus with $N,Z \gtrsim 50$.  We also note the
unusual behavior of the $^{78}$Ni results for $\eMax/\EMax=14/14$ at
$\hbar\Omega=28\MeV$ in Fig.~\ref{fig:NiEgsConvergence} is probably
caused by truncation artifacts due to the $\EMax$ cut.

In Fig.~\ref{fig:NiRchConvergence}, we show the model-space
convergence of the charge radii for $^{56}$Ni and $^{78}$Ni.  Similar
to the calcium isotopes discussed above, we see a gradual increase
with increasing SRG resolution scale and a larger value for the
2.0/2.0 (PWA) interaction.  While the results for $^{56}$Ni appear
well converged for all starting Hamiltonians, this is less the case
for $^{78}$Ni either with respect to $\eMax$ or $\EMax$, and
especially for the larger cutoffs and the 2.0/2.0 (PWA) interaction.

Finally, before studying the systematics of the ground-state energies
and charge radii of closed shell nuclei, we compare our results to the
coupled-cluster calculations of Hagen \etal~\cite{Hage16Ni78} for the
1.8/2.0 (EM) interaction. Considering the same model-space truncation
$\eMax/\EMax=14/16$ and harmonic-oscillator frequency
$\hbar\Omega=16\MeV$ we find good agreement within $\approx 1\%$ for
$^{16}$O: $-127.2\MeV$ (IM-SRG(2)) vs.~$-128\MeV$ ($\Lambda$-CCSD(T));
for $^{40}$Ca: $-344.5\MeV$ vs.~$-348\MeV$; for $^{48}$Ca:
$-416.1\MeV$ vs.~$-419\MeV$; and for $^{78}$Ni: $-633.6\MeV$
vs.~$-637\MeV$, while there is a difference of more than $3\%$ for
$^{4}$He ($-29.2\MeV$ vs.~$-28.2\MeV$).

\begin{figure}[t]
\includegraphics[width=\columnwidth]{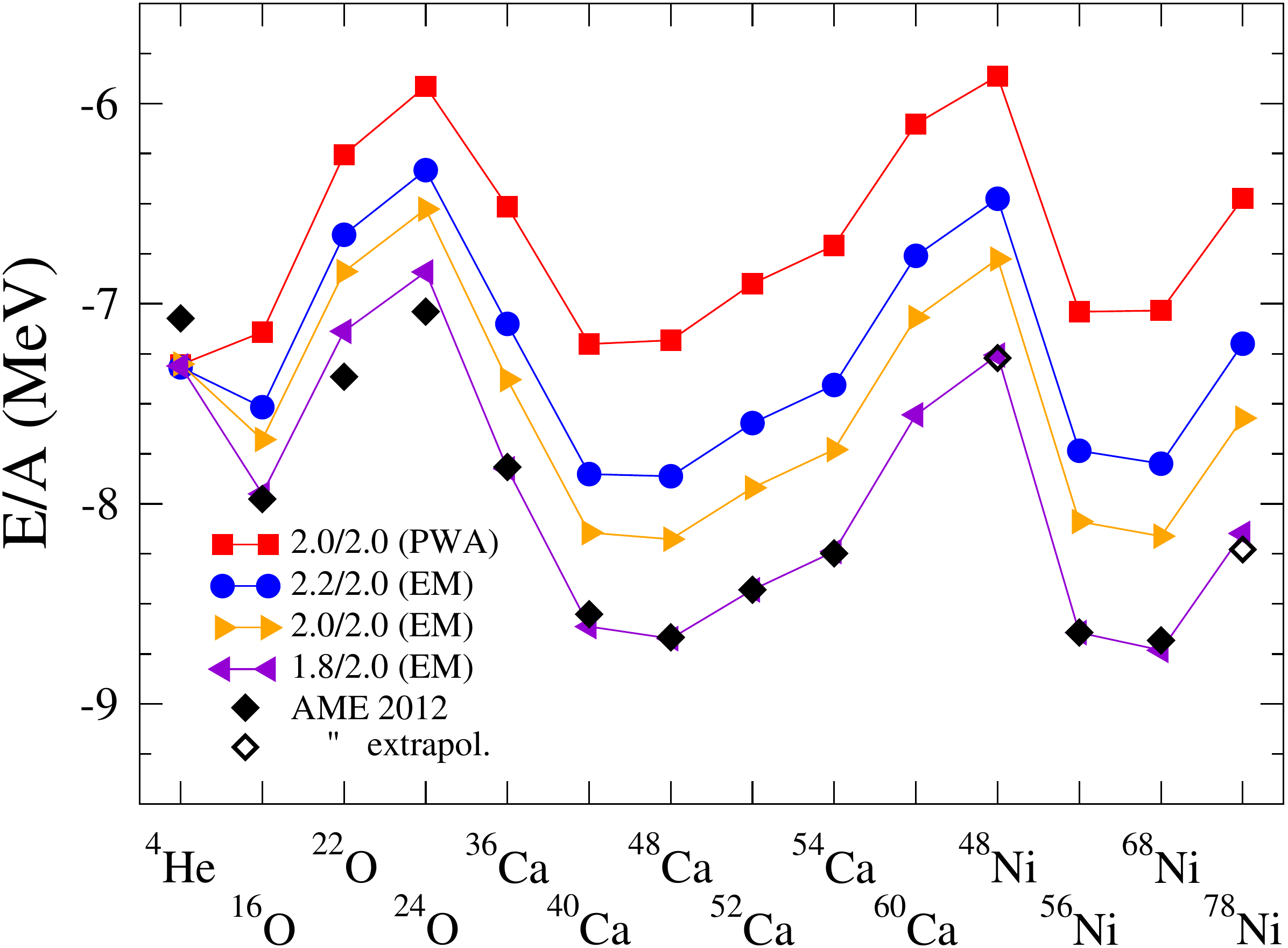}
\caption{Systematics of the energy per nucleon $E/A$ of closed-shell 
nuclei from $^{4}$He to $^{78}$Ni calculated with the IM-SRG for
the four Hamiltonians considered. The results are compared against
experimental ground-state energies from the AME 2012~\cite{Wang12AME12}
(extrapolated for $^{48, 78}$Ni).}
\label{fig:EgsdivA}
\end{figure}

\begin{figure}[t]
\includegraphics[width=\columnwidth]{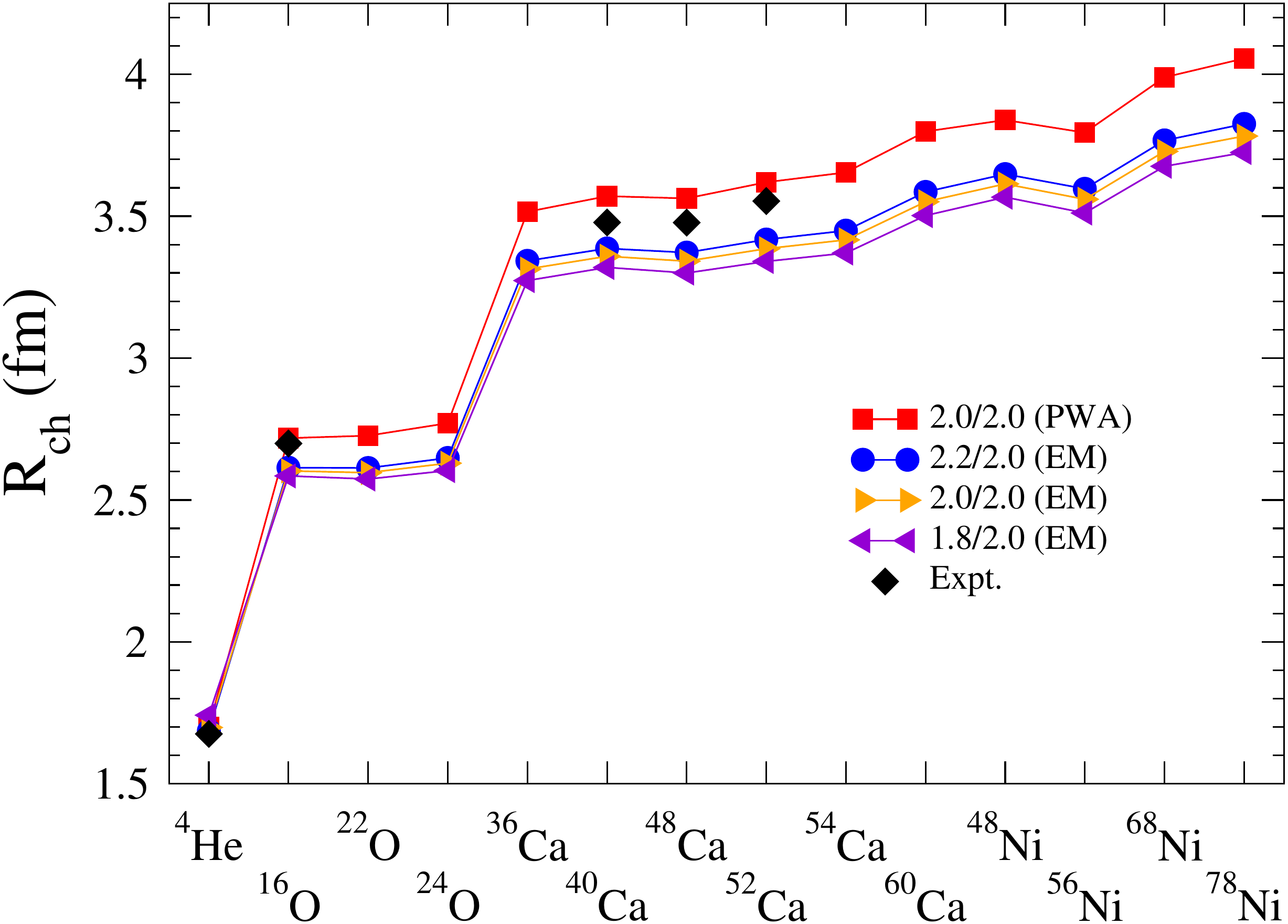}
\caption{Systematics of charge radii of closed-shell nuclei from $^{4}$He 
to $^{78}$Ni calculated with the IM-SRG for the four Hamiltonians considered.
The results are compared against experimental charge radii~\cite{Ange13rch}
where available.}
\label{fig:Rch}
\end{figure}

Finally, in Figs.~\ref{fig:EgsdivA} and \ref{fig:Rch} we show
ground-state energies and charge radii, respectively, for selected
closed-shell nuclei from $^{4}$He to $^{78}$Ni. Except for the
neutron-rich oxygen isotopes $^{22,24}$O all calculated ground-state
energies from the 1.8/2.0 (EM) interaction are in very good agreement
with experiment. Interestingly the other three interactions follow the
same pattern but are shifted by as much as $1.5\MeVdivA$ in the case
of the 2.0/2.0 (PWA) interaction.  The experimental charge radii are
enclosed by the 2.2/2.0 (EM) and 2.0/2.0 (PWA) results, but the trend
observed for the closed-shell nuclei studied in detail already above
appears to hold at least up to $^{78}$Ni. That is, radii with
1.8-2.2/2.0 are too small, but 2.0/2.0 (PWA) gives slightly too large
radii.  As in the case of ground-state energies, the radii systematics
is similar for all Hamiltonians; with mainly only a constant shift for
the different interactions. This behavior for the ground-state energy
and charge radii is reminiscent of the Coester-like line for the
saturation points of the four Hamiltonians
considered~\cite{Dris16asym}.

\section{Open-shell isotopic chains}
\label{sec:Chains} 

In this section, we move beyond closed-shell systems to explore
ground- and excited-state systematics throughout a selection of
isotopic chains in the $sd$ and $pf$ shells, namely sodium, sulfur,
calcium, manganese, and nickel. The VS-IM-SRG method used here was
shown to agree with large-space methods to better than 1\% for
ground-state energies~\cite{Stro17ENO}.

We also calculate charge radii, less studied within the context of ab
initio
approaches~\cite{Cipo13Ox,Ekst15sat,Hage16NatPhys,Ruiz16Calcium,Lapo16radiiO},
with the VS-IM-SRG for the first time, where the proton mean-square
radius operator of Eq.~\eqref{eq:def_rp} is transformed via the same
unitary transformation as the Hamiltonian. This gives a valence-space
radius operator to be used with valence-space wave functions, after
which the core point-proton radius and corrections of
Eq.~\eqref{eq:def_rch} are applied to obtain the absolute charge
radius. We note that induced two-body corrections to the radius
operator are included naturally in the VS-IM-SRG formalism.

\subsection{Ground-state energies and radii}

Given the remarkable agreement with experimental ground-state energies
for closed-shell systems from the 1.8/2.0 (EM) interaction discussed
in Sec.~\ref{sec:ClosedShell}, we compare the systematics of
ground-state energies calculated with this interaction to experimental
data where they exist in the isotopic chains mentioned above.  For
sodium, sulfur, and calcium we take $\eMax/\EMax=12/16$, while for
manganese and nickel we use $\eMax/\EMax=14/16$. In all cases
$\hbar\Omega=16\MeV$ is taken for the harmonic-oscillator frequency.
In addition, we directly compare single-reference IM-SRG and
valence-space results in calcium and nickel for the closed-shell
cases.  The valence space is defined to be one major
harmonic-oscillator shell for protons and neutrons.  For example, for
the sulfur chain we take a proton and neutron $sd$ valence space above
an $^{16}$O core for $N<20$, a proton $sd$ valence space above a
$^{28}$O core for $N=20$, and a proton $sd$ neutron $pf$ valence space
above a $^{28}$O core for $N>20$.  It should
be noted that at oscillator shell closures for neutrons,
no explicit neutron excitations are allowed in the valence
space. In particular for systems near
the transition from one valence space to another, con-
tributions from cross-shell excitations will be important.
These excitations are incorporated approximately 
by the IM-SRG decoupling, and our truncation to
two-body operators is insufficient for these isotopes.
We mark these oscillator closures as a vertical
dotted line in all figures.
While preliminary efforts indicate that the VS-IM-SRG
approach is capable of decoupling the relevant mixed va-
lence spaces, and thus treating these excitations explicitly,
we will address this issue in a future work.

\begin{figure}[t]
\centering
\includegraphics[width=0.9\columnwidth]{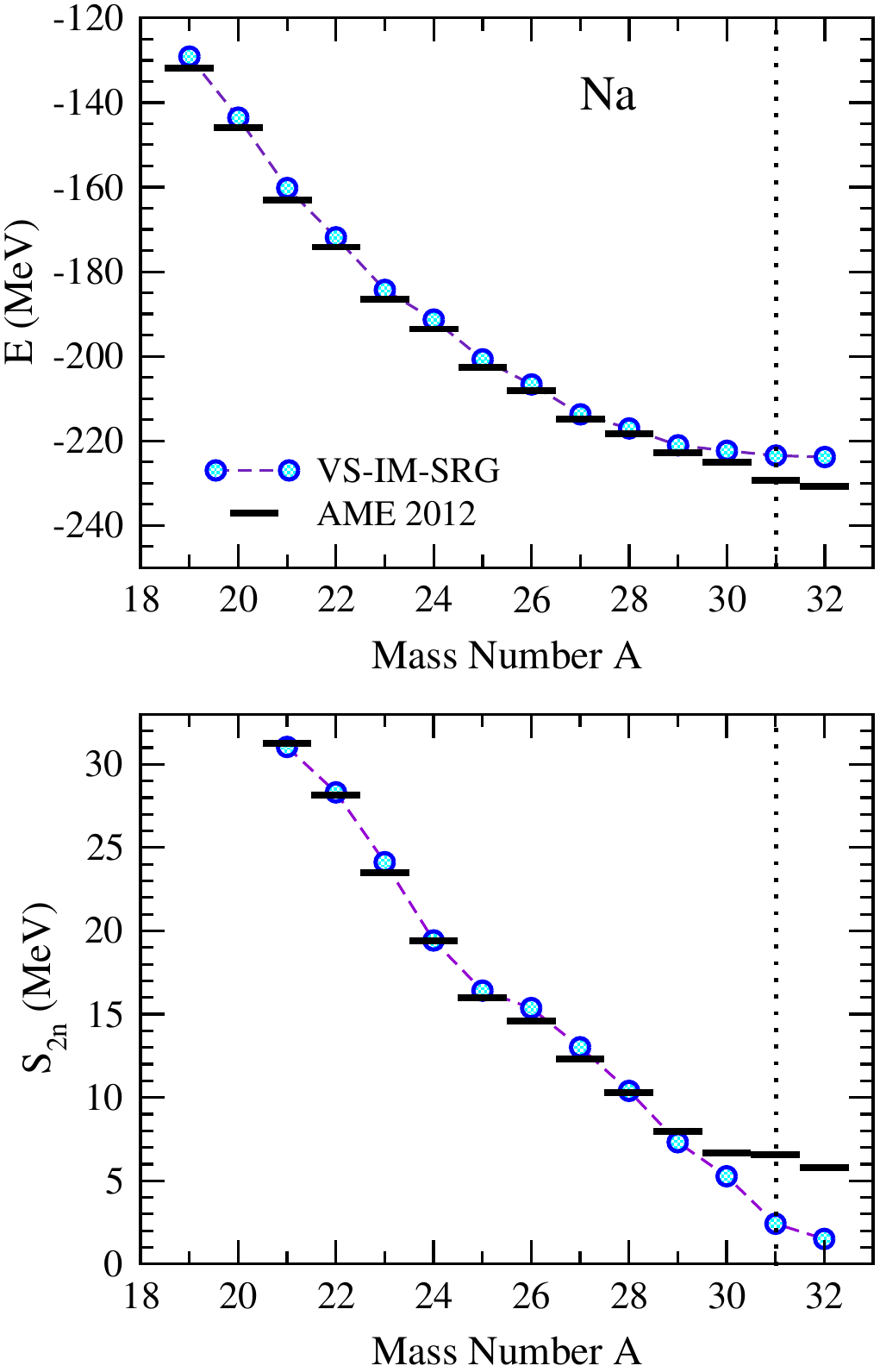}
\caption{Ground-state energies (top) and two-neutron separation energies
(bottom) of sodium isotopes for the 1.8/2.0 (EM) Hamiltonian (circles) 
compared to experiment (AME 2012, bars)~\cite{Wang12AME12}. See text for
details on the valence spaces used. The vertical dotted line marks the end of
the $sd$ shell at $N=20$.}
\label{fig:Nags}
\end{figure}

\begin{figure}[t]
\centering
\includegraphics[width=0.9\columnwidth]{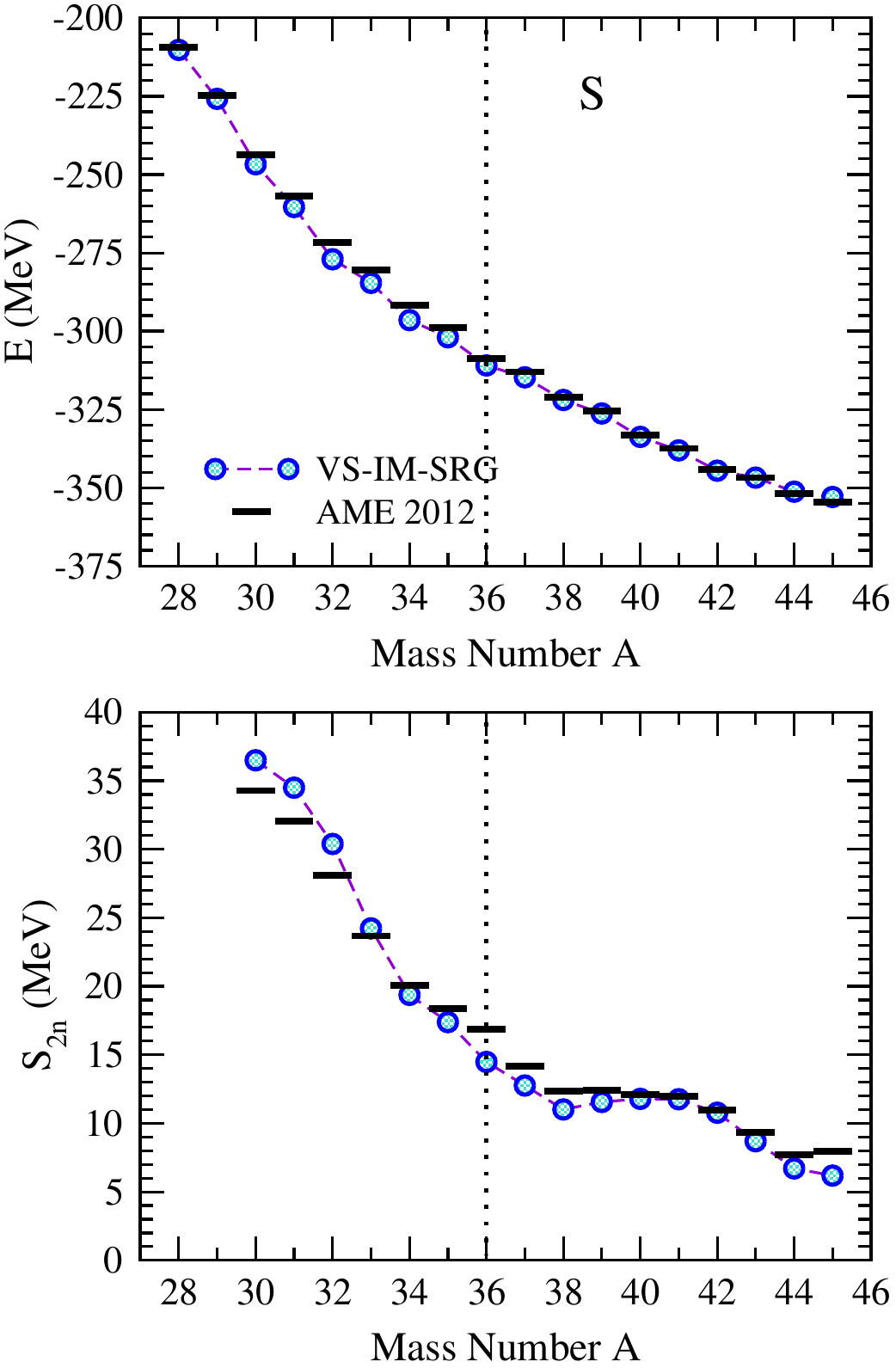}
\caption{Ground-state energies (top) and two-neutron separation energies
(bottom) of sulfur isotopes compared to experiment, with the same legend
and details as in Fig.~\ref{fig:Nags}.}
\label{fig:Sgs}
\end{figure}

Beginning in the $sd$ region, we show in Figs.~\ref{fig:Nags}
and~\ref{fig:Sgs}, ground-state energies and two-neutron separation
energies $\Sn$ for sodium and sulfur isotopes, respectively.  In both
cases we find good agreement with absolute experimental ground-state
energies, outside of $^{31,32}$Na, which are somewhat underbound.  The
ground states of $^{30-32}$Na are dominated by deformed
configurations~\cite{Trip07na30,Door10na31}, not captured in neutron
$sd$ or $pf$ valence-space calculations.  These island-of-inversion
isotopes will be investigated further in the context of decoupling
neutron $sd-pf$ cross-shell valence spaces.  Likewise, the $\Sn$
results are in remarkable agreement with data, except in the
region near $N=20$.  In Ref.~\cite{Stro17ENO}, sodium
isotopes were also investigated with the VS-IM-SRG approach, but
instead using the EM $500\MeV$ potential with local N$^2$LO 3N
forces~\cite{Navr07local3N} consistently SRG evolved to $\lambda =
1.88\fmi$.  For this choice of Hamiltonian the isotopes $^{22}$Na up
to $^{32}$Na are overbound, while the rest of the chain is in good
agreement with experiment.  With three protons above the closed $Z=8$
proton shell, no other ab initio method is currently able to calculate
sodium isotopes.  Except for the single-reference and VS-IM-SRG calculations of
$^{32,36}$S with the SRG-evolved NN+3N forces mentioned above~\cite{Stro17ENO}, which
display significant overbinding not seen with the 1.8/2.0 (EM) interaction
used here, there are no other ab initio calculations available for these
open-shell sulfur isotopes.

\begin{figure}[t]
\centering
\includegraphics[width=0.9\columnwidth]{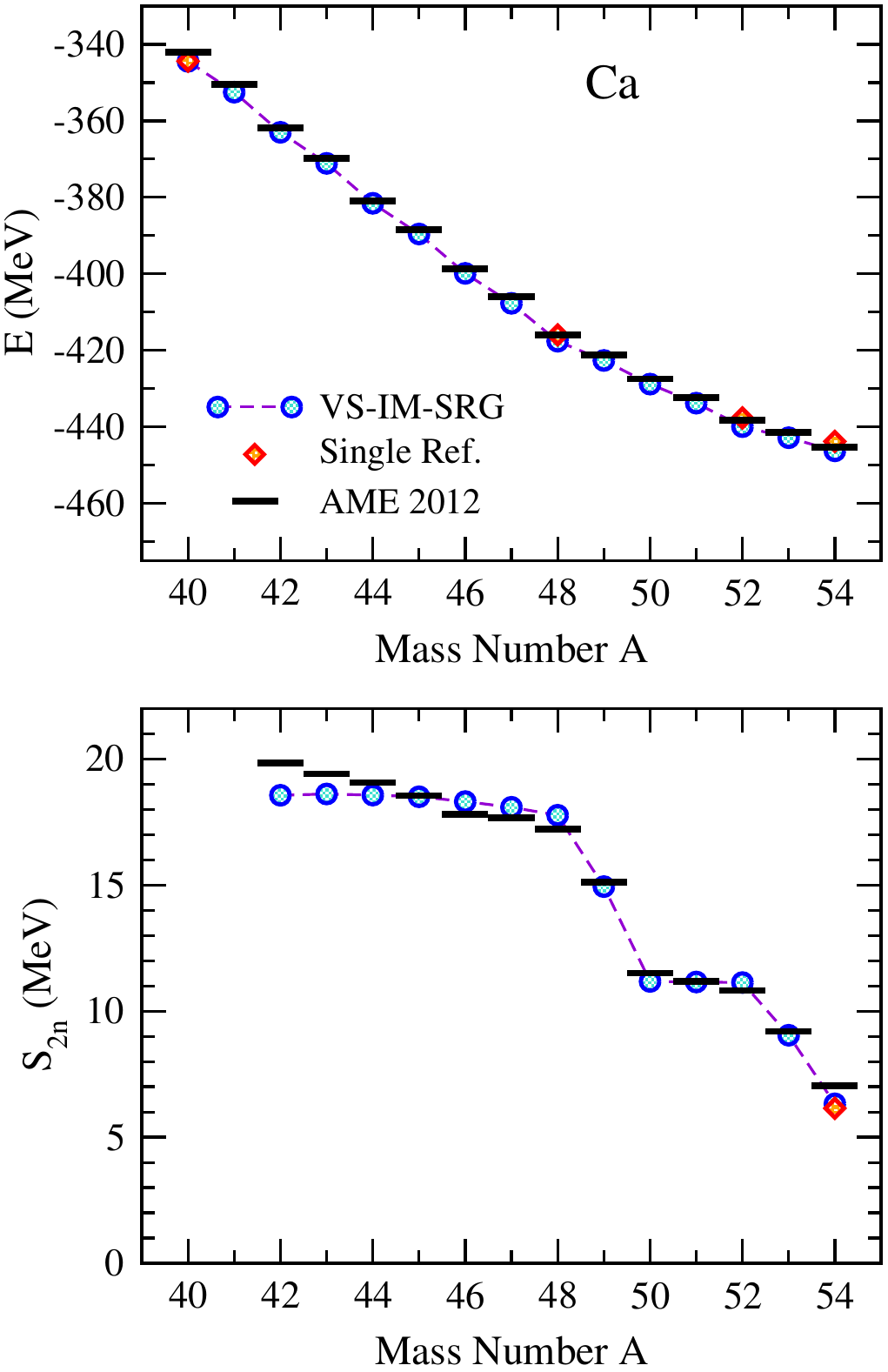}
\caption{Ground-state energies (top) and two-neutron separation energies
(bottom) of calcium isotopes for the 1.8/2.0 (EM) Hamiltonian (circles) 
compared to experiment (AME 2012, bars)~\cite{Wang12AME12}. See text for
details on the valence spaces used. For closed-subshell isotopes we 
also show the results of the single-reference IM-SRG (diamonds) for 
comparison.}
\label{fig:Cags}
\end{figure}

\begin{figure}[t]
\centering
\includegraphics[width=0.9\columnwidth]{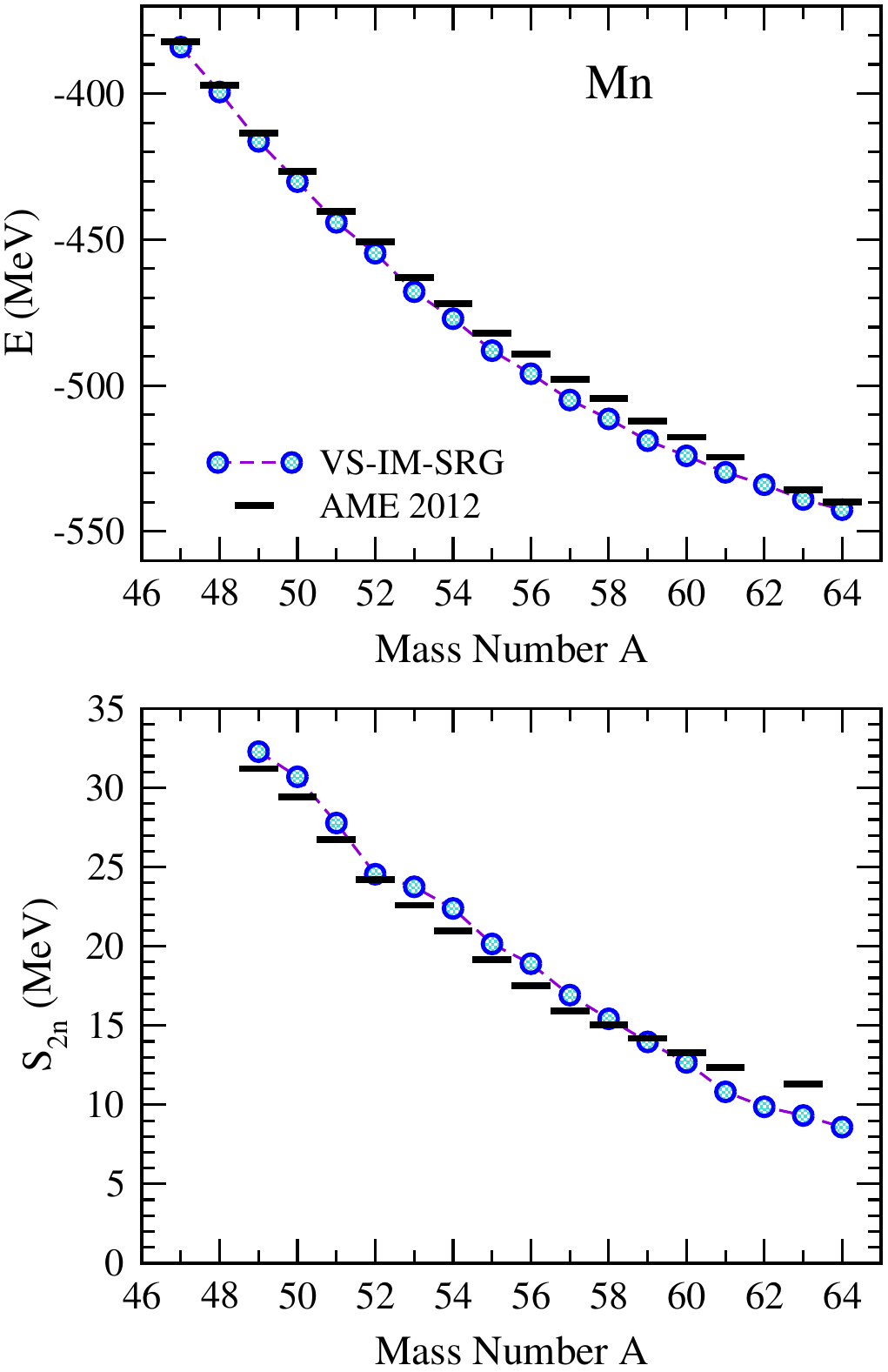}
\caption{Ground-state energies (top) and two-neutron separation energies
(bottom) of manganese isotopes compared to experiment, with the same legend
and details as in Fig.~\ref{fig:Cags}.}
\label{fig:Mngs}
\end{figure}

\begin{figure}[t]
\centering
\includegraphics[width=0.9\columnwidth]{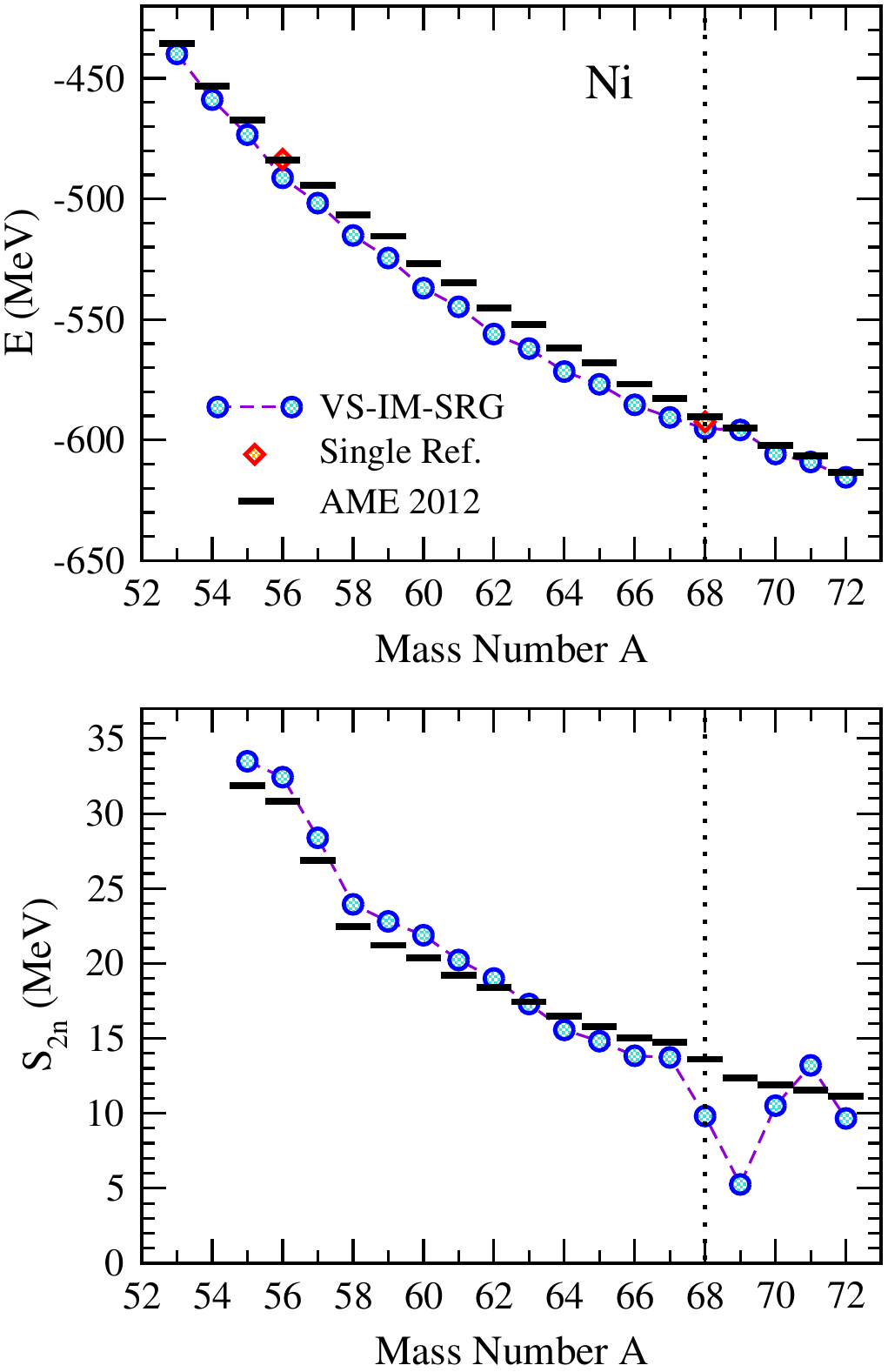}
\caption{Ground-state energies (top) and two-neutron separation energies
(bottom) of nickel isotopes compared to experiment, with the same legend
and details as in Fig.~\ref{fig:Cags}. The vertical dotted line marks 
the end of the $pf$ shell at $N=40$.}
\label{fig:Nigs}
\end{figure}

In the $pf$ shell, the agreement with experimental data remains good
as well, as shown in Figs.~\ref{fig:Cags}, \ref{fig:Mngs},
and~\ref{fig:Nigs} for calcium, manganese, and nickel isotopic chains,
respectively.  In calcium, we also compare with the corresponding
single-reference results for $^{40,48,52,54}$Ca, where, as noted in
Ref.~\cite{Stro17ENO}, the VS-IM-SRG results agree with the
single-reference calculations to better than 1\%.  We also reproduce
well the sharp decreases in $\Sn$ values after $N=28$ and $N=32$, in
good agreement with recent precision
experiments~\cite{Gall12Ca,Wien13Nat}, indicating that the shell
closures at $N=28$ and $N=32$ are well reproduced with the 1.8/2.0
(EM) interaction. We therefore also expect predictions of $\Sn$ values
past $N=34$ to be reliable, at least qualitatively, when the data
become available. Similar good agreement is seen for all trends in the
manganese isotopes, which, with five protons above the $Z=20$ proton
shell closure, are currently inaccessible to all other ab initio
methods.  Finally, we see that throughout the nickel chain, absolute
ground-state energies become modestly overbound in the mid-shell
region on the order of up to 10$\MeV$. All other
experimental trends (aside from the artificial kink in the vicinity of
$N=40$) are well reproduced, including the
sharp drop past $N=28$. We
also note that the somewhat larger discrepancy between
single-reference and VS-IM-SRG results for $^{56}$Ni is likely due to
the ground-state configuration obtained in the valence-space
diagonalization being only 30\% pure filled proton and neutron
$f_{7/2}$. While experimental energies are known past $A=72$, this
represents our current limitation of diagonalizing the valence-space
Hamiltonian exactly with modest computational resources.  Using
standard extensions and/or controlled truncations, isotopes as heavy
as $^{80}$Ni may be reached, though as seen in
Sec.~\ref{sec:ClosedShell}, such results may not be completely
converged in terms of $\EMax$. The results for the calcium and nickel
isotopes using the consistently SRG-evolved EM $500\MeV$ potential
with local N$^2$LO 3N forces~\cite{Navr07local3N} are significantly
overbound up to 100$\MeV$~\cite{Stro17ENO}, highlighting the
importance of considering saturation for whether chiral interactions
can describe bulk properties of nuclei across the nuclear chart.

\begin{figure}[t]
\centering
\includegraphics[width=0.9\columnwidth]{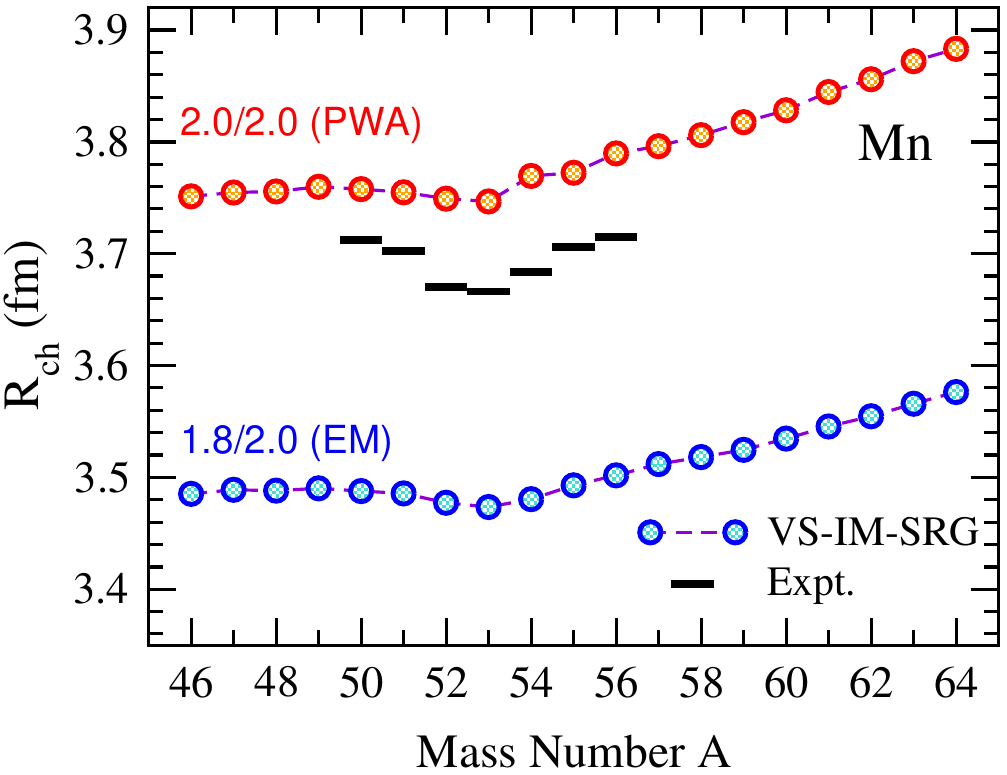}
\caption{Charge radii of manganese isotopes for the 1.8/2.0 (EM) (blue circles)
and 2.0/2.0 (PWA) (red circles) Hamiltonian compared to experiment~\cite{Ange13rch}.
See text for details on the valence spaces used.}
\label{fig:Mnrad}
\end{figure}

While we reserve a complete discussion of charge radii systematics
within the VS-IM-SRG for a future work, we illustrate the versatility
of this approach in Fig.~\ref{fig:Mnrad}, where we plot radii for the
complete $pf$-shell manganese chain, comparing with experimental data
where available.  From Fig.~\ref{fig:Rch}, we expect charge radii predicted
with the 1.8/2.0 (EM) and 2.0/2.0 (PWA) interactions to be systematically too
small and too large, respectively. While this is indeed seen, an interesting trend
in charge radii is predicted in both cases, with a roughly parabolic shape to $N=28$,
followed by a sharp increase for $N>28$.  Experimental data are quite limited
in manganese isotopes, but available data do seem to show this trend,
albeit with more pronounced structures, as also seen in recent
experimental measurements of charge radii in calcium
isotopes~\cite{Ruiz16Calcium}.  While neither interaction perfectly reproduces
experiment, 2.0/2.0 (PWA) only moderately overpredicts charge radii  and
should provide a reasonably reliable guide to trends across isotopic chains.
The general absence of systematic data highlights the importance of continued
systematic experimental investigations of charge radii.

\begin{figure}[t]
\centering
\includegraphics[width=0.9\columnwidth]{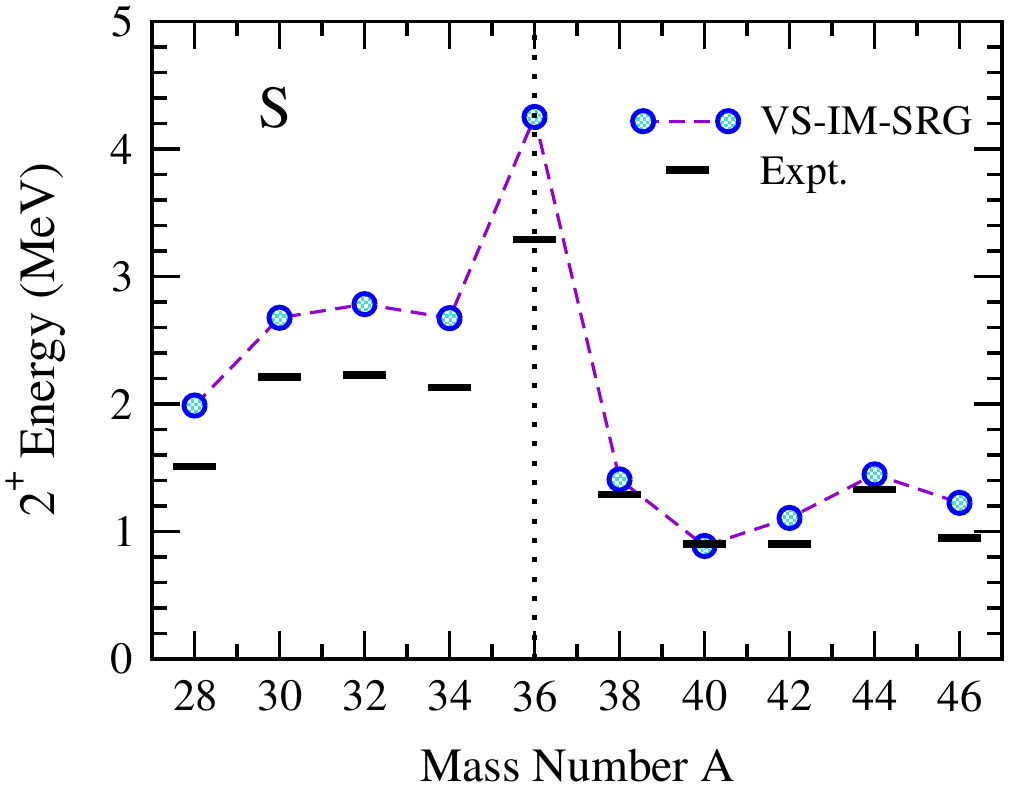}
\caption{First excited 2$^+$ energies of even sulfur isotopes for the 
1.8/2.0 (EM) Hamiltonian (circles) compared to
experiment~\cite{nndc14ENSDF}. See text for details on the valence 
spaces used. The vertical dotted line marks the end of the $sd$ shell
at $N=20$.}
\label{fig:S2+}
\end{figure}

\subsection{Excited states}

Given the remarkable description of experimental ground-state
properties from the 1.8/2.0 (EM) interaction, it is also of interest
to investigate to what extent the structure of excited states is
captured.  In the VS-IM-SRG approach, all excited states allowed
within a given valence space are obtained directly via
diagonalization.  Here we focus on first excited $2^+$ states and
associated shell closures in the subset of even-even sulfur, calcium,
and nickel isotopes.

Beginning with sulfur, shown in Fig.~\ref{fig:S2+}, we see an overall
good reproduction of the experimental trends in $2^+$ energies.
When neutrons occupy the $sd$
valence space, however, these energies are systematically several
hundred keV too high. Beyond $N=20$, when the neutron valence space
changes to the $pf$ shell, agreement with data improves,
including the modest peak at $N=28$ in $^{44}$S.  Given the absence of
allowed neutron excitations at $N=20$, the $2^+$ energy here is
artificially too high and is expected to decrease when such degrees of
freedom are included in the valence space.

\begin{figure}[t]
\centering
\includegraphics[width=0.9\columnwidth]{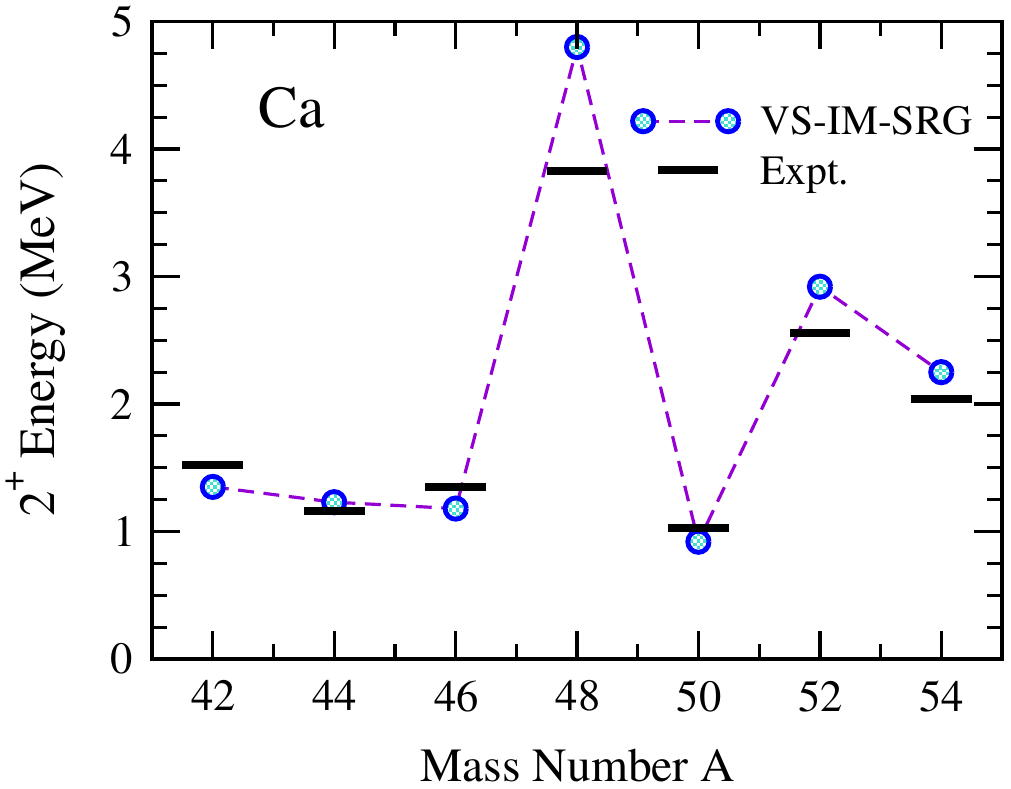}
\caption{First excited 2$^+$ energies of even calcium isotopes for the
1.8/2.0 (EM) Hamiltonian (circles) compared to experiment~\cite{nndc14ENSDF}.
See text for details on the valence spaces used.}
\label{fig:Ca2+}
\end{figure}

\begin{figure}[t]
\centering
\includegraphics[width=0.9\columnwidth]{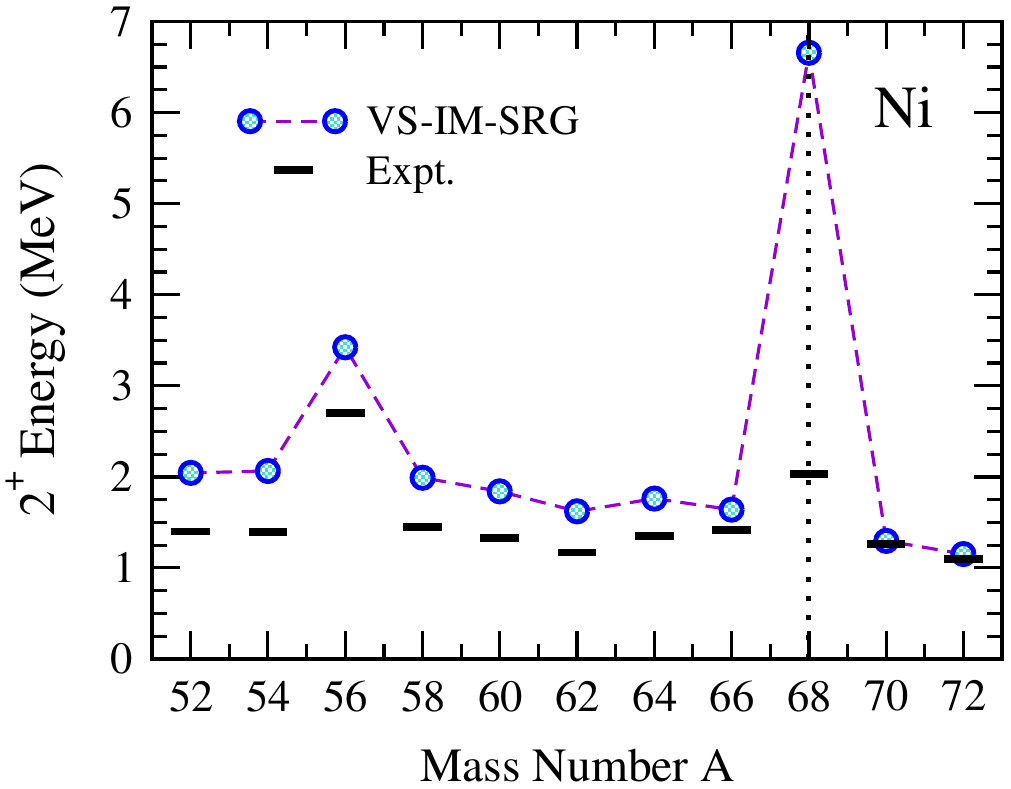}
\caption{First excited 2$^+$ energies of even nickel isotopes for the
1.8/2.0 (EM) Hamiltonian (circles) compared to
experiment~\cite{nndc14ENSDF}.  See text for details on the valence
spaces used.  The vertical dotted line marks the end of the $pf$ shell
at $N=40$.}
\label{fig:Ni2+}
\end{figure}

For the calcium isotopes, shown in Fig.~\ref{fig:Ca2+}, the calculated
results agree well with data for open-shell cases.  While
relative peaks are seen at the $N=28,32$ shell closures as well as the
recently measured $N=34$ closure in $^{54}$Ca~\cite{Step13Ca54}, they
are systematically too high, particularly in $^{48}$Ca.  While we
might initially attribute this to neglected proton excitations due to
the choice of valence space, similar features are also seen in the
nickel isotopes, which allow both proton and neutron excitations
except at $N=40$.  For the nickel isotopes, a similar picture to
sulfur is seen in Fig.~\ref{fig:Ni2+}. Where when neutrons fill the
$pf$ shell, the $2^+$ energies reproduce the experimental trend, but
are systematically several hundred keV too high.  When neutrons begin
filling the $sdg$ orbits past $^{68}$Ni, the results agree very well
with data, making predictions out to $^{80}$Ni possible to investigate
the closed-shell nature of $^{78}$Ni.  The very high $2^+$ state in
$^{68}$Ni is clearly due to a lack of allowed neutron excitations, and
should again be as an artifact of the many-body approximation.

The pattern of too-high $2^+$ energies in closed-shell systems is a
common feature of our calculations.  The origin of this behavior is
unclear, but some direction might be provided by coupled-cluster
calculations of $^{48}$Ca and $^{78}$Ni~\cite{Hage16Ni78}. In this
work, it was found that using the same 1.8/2.0 (EM) interaction, when
particle-hole excitations were limited to the coupled-cluster singles
and doubles (CCSD) approximation, the first excited $2^+$ state in
$^{48}$Ca was approximately $1\MeV$ too high, very close to our
result.  When perturbative triples excitations were then included,
this energy was lowered to close to the experimental value, and a
similar decrease was seen in the first $2^+$ energy in
$^{78}$Ni. Because the IM-SRG(2) approximation in this work is
analogous to the CCSD truncation~\cite{Herg16PR}, we might expect a
similar improvement in extensions of IM-SRG(2) analogous to the
perturbative triples of coupled cluster.  Due to the final step of
diagonalizing in the valence space, however, the expectation would be
that much of this physics should already be captured, as long as
excitations near the Fermi surface were of dominant importance.  As
the development of a controlled approximation to IM-SRG(3) is
currently in progress, we so far have no means to check whether such
an improvement will remedy the too-high $2^+$ states at closed shells.
Nevertheless, this appears to be a deficiency in the many-body method,
not the interaction.

\section{Summary and conclusions}

We have performed ab initio IM-SRG calculations of ground-state
energies and charge radii of broad range of closed- and open-shell
nuclei with $A \leqslant 78$. For radii of open-shell nuclei, these
represent first valence-space IM-SRG results. We have focused on a set
of chiral low-resolution NN+3N interactions that predict realistic
saturation properties. As a baseline, we have first studied the
convergence properties of these Hamiltonians with respect to
model-space truncations for both two- and three-body interactions.
Each of the NN+3N interactions used here reproduces few-body data with
equivalent accuracy.  However, the interactions do not produce
equivalent results for medium-mass nuclei. In fact, the systematics of
ground-state energies and radii indicates that the difference is
dominantly due to their different nuclear matter saturation properties.

One particular interaction yields energies in good agreement with
experiment from light nuclei up to the point at which we are limited
by the convergence of the many-body calculation.  This behavior
appears to be accidental, in the sense that we could not have
anticipated from the few-body results which of these interactions
would produce the desired absolute energies, but it suggests two
conclusions.  First, that the operator structures contained in these
chiral interactions (NN at N$^{3}$LO and 3N at N$^{2}$LO) are
sufficient to describe many of the features of the energies of light
and medium-mass nuclei, while future consistent calculations at
N$^3$LO (and N$^2$LO) are of course called for. Second, as suggested
in Ref.~\cite{Ekst15sat}, saturation properties are essential for this
accurate description. Both of these points highlight the importance of
nuclear matter as a theoretical benchmark and guidance for the
development of next-generation chiral interactions.

\begin{acknowledgments}
This work was supported by the ERC Grant No.~307986 STRONGINT,
the DFG under Grant SFB 1245, the BMBF under Contract No.~05P15RDFN1,
the National Research Council of Canada and NSERC. Computations were
performed with an allocation of computing resources at the J\"ulich
Supercomputing Center (JURECA), at the Computing Center of the TU
Darmstadt (Lichtenberg), and at the Max-Planck-Institute for Nuclear
Physics.
\end{acknowledgments}

\bibliographystyle{apsrev4-1}
\bibliography{strongint}

\end{document}